\documentclass[9pt,twocolumn,twoside]{osajnl}
\usepackage{amsmath,amsfonts,graphicx}
\usepackage{subfig}
\usepackage{siunitx}
\usepackage[flushleft]{threeparttable}
\usepackage{stackengine}
\usepackage{soul}
\DeclareMathOperator{\sech}{sech}
\journal{ol}
\setboolean{shortarticle}{true}
\title{Tailoring $\mathcal{PT}$-symmetric soliton switch}
\author[1,*]{A. Govindarajan}
\author[2]{Amarendra K. Sarma}
\author[1]{M. Lakshmanan}
\affil[1]{Centre for Nonlinear Dynamics, School of Physics, Bharathidasan University, Tiruchirappalli - 620 024, India}
\affil[2]{Department of Physics, Indian Institute of Technology Guwahati, Guwahati, 781039, Assam, India}
\affil[*]{Corresponding author: govin.nld@gmail.com}
 \dates{}
\ociscodes{(060.1810)   Buffers, couplers, routers, switches, and multiplexers; (190.7110)   Ultrafast nonlinear optics; (250.6715)   All-optical devices.}
\doi{\url{}}
\begin{abstract}
We theoretically demonstrate soliton steering in $\mathcal{PT}$-symmetric coupled nonlinear dimers. We show that if the length of the $\mathcal{PT}$-symmetric system is set to $2\pi$ contrary to the conventional one which operates satisfactorily well only at the half-beat coupling length, the $\mathcal{PT}$ dimer remarkably yields an ideal soliton switch exhibiting almost 99.99\% energy efficiency with an ultra-low critical power.
\end{abstract}
\setboolean{displaycopyright}{true}
\begin{document}
\maketitle
Since the first observation of both parity ($\mathcal{P}$) and time ($\mathcal{T}$) symmetry in a coupled waveguide with gain in one core and equal amount of loss in the other core \cite{ruter2010observation}, $\mathcal{PT}$-symmetric coupled dimers have been found to act as one of the most fertile sources for testing various $\mathcal{PT}$ related notions and concepts \cite{El-Ganainy2018}. Moreover, such systems exhibit many unusual exotic phenomena like  non-reciprocity of pulse propagation and power oscillations. Subsequently, the impact of $\mathcal{PT}$-symmetry was explored in a host of  diverse $\mathcal{PT}$-symmetric optical systems, which include periodic structures \cite{musslimani2008}, micro-ring resonator \cite{hodaei2014}, laser absorber \cite{SLonghi}, fiber Bragg grating \cite{phang2014impact} and metamaterials \cite{feng2013}.  They also exhibit other interesting dynamics like Bloch oscillations, unidirectional invisibility, etc. \cite{lin2011uni,suchkov2016nonlinear, PRA.94.023829}. Most of the aforementioned works are generally devoted to study the different roles of $\mathcal{PT}$-symmetry on the fiber couplers rather than on the steering dynamics, which is an inevitable and a fundamental application of fiber couplers, influenced by the $\mathcal{PT}$ symmetric effect \cite{ramaswami2009}. As a matter of fact, all-optical switching  is a vital and central component of the integrated devices in lightwave communication systems, which needs  specific attention in  $\mathcal{PT}$-symmetric couplers. Though an earlier study emphasizing the continuous wave (CW) switch has been carried out many years ago in a system of gain and loss \cite{chen1992}, it is a well established fact that the CW is a very poor candidate as it tends to break into multiple pieces during the nonlinear switching and thus one should consider the soliton pulse \cite{trillo_switch,kivshar1993switching} due to its particle like nature and also as a prominent one  among other types of pulses including Gaussian and super-Gaussian. The objective of the present Letter is to explore the steering dynamics of optical solitons in nonlinear directional couplers influenced by such $\mathcal{PT}$-symmetric effect. It is important to note that stability of solitons in such a $\mathcal{PT}$-symmetric dimer has recently been studied by Driben and Malomed \cite{driben2011}. As regards soliton switch, the practical implementation of all-optical soliton switching is put in dormant for many years due to the requirement of high amount of critical power ($P_{cr}$) at which equal sharing of energy takes place between the parallel (bar) and the crossed states \cite{Agrawal2001}.

In the present work we show that the need of huge amount of critical power is drastically reduced in  couplers by exploiting the $\mathcal{PT}$-symmetry through the combination of equal amount of gain and loss in their two arms.
\begin{figure}[t]
	\centering
	\subfloat[Type 1 $\mathcal{PT}$-symmetric coupler]{\includegraphics[width=0.5\linewidth]{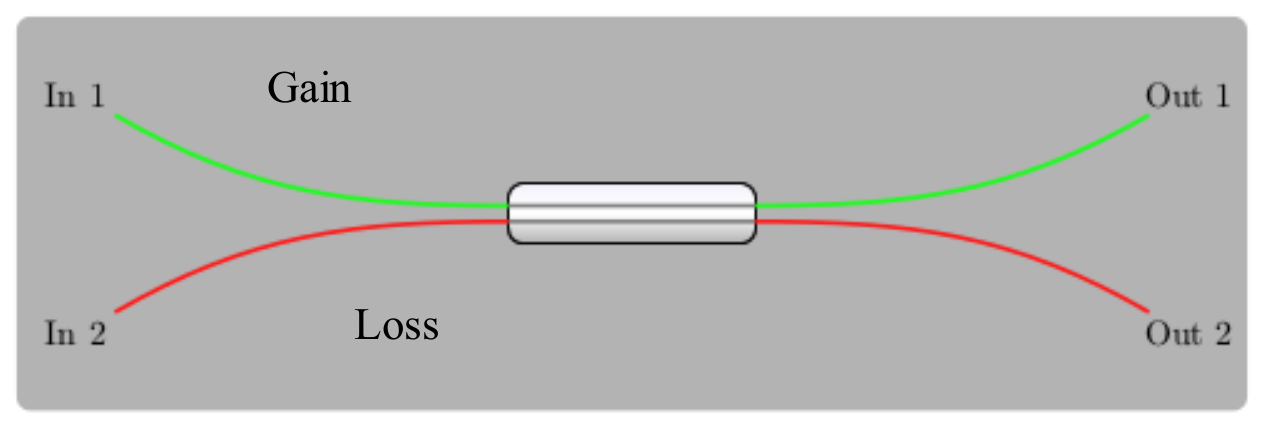}}
	\subfloat[Type 2 $\mathcal{PT}$-symmetric coupler]{\includegraphics[width=0.5\linewidth]{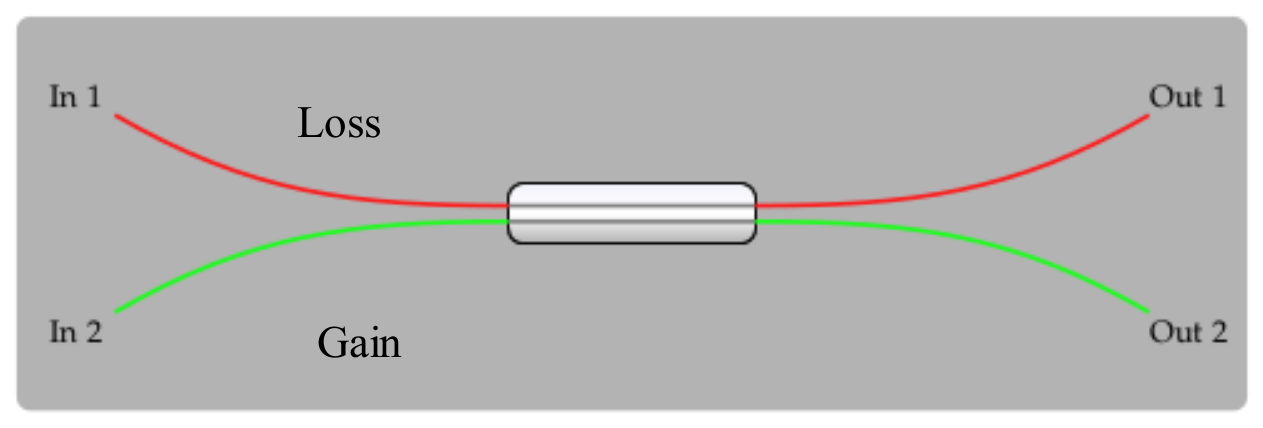}}
	\vfill
	\subfloat[Steering dynamics of solitons in $\mathcal{PT}$-symmetric dimers ]{\includegraphics[width=0.8\linewidth]{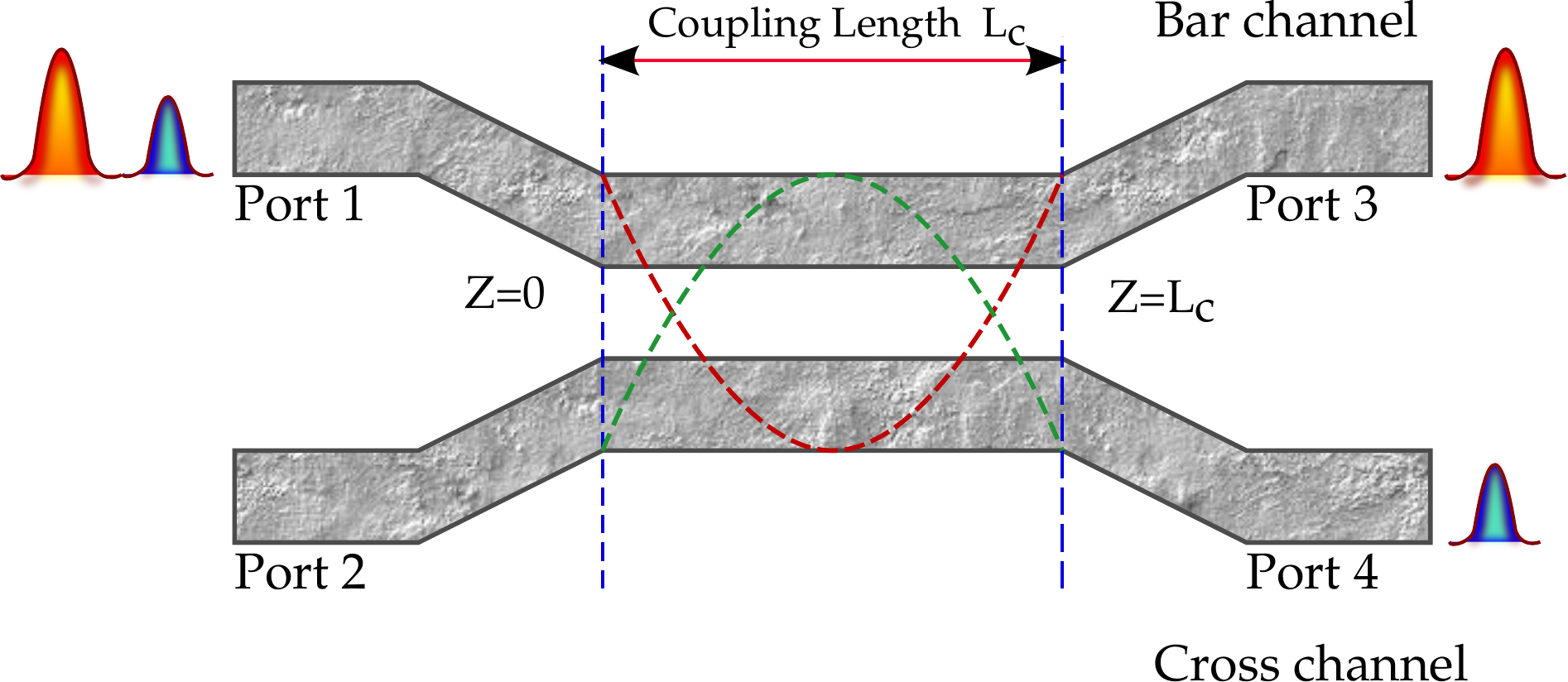}}
	\caption{Schematic illustrations showing the different configurations of $\mathcal{PT}$-symmetries and a functional diagram of the system under study.}
	\label{FigSchematic}
\end{figure}

The dynamics of solitons in $\mathcal{PT}$-symmetric dimers (see Fig. \ref{FigSchematic}) can be described by a system of coupled nonlinear Schr\"{o}dinger equations. In such $\mathcal{PT}$-symmetric dimers,  the optical pulse propagation with balanced gain and loss is based on the slowly varying envelope (paraxial) approximation (SVEA) and the evolution equation with the so-called retarded time $\tau$ and the propagation distance $\zeta$ is written in the scaled form as \cite{driben2011}
\begin{gather}
i\frac{\partial \Psi_{1}}{\partial \zeta }+\frac{1}{2}\frac{\partial
	^{2}\Psi_{1}}{\partial \tau ^{2}}+\left\vert \Psi_{1}\right\vert ^{2}\Psi_{1}+\kappa
\Psi_{2}=i\Gamma \Psi_{1},  \label{eqn:1} \\
i\frac{\partial \Psi_{2}}{\partial \zeta }+\frac{1}{2}\frac{\partial
	^{2}\Psi_{2}}{\partial \tau ^{2}}+\left\vert \Psi_{2}\right\vert ^{2}\Psi_{2}+\kappa
\Psi_{1}=-i\Gamma \Psi_{2},
\label{eqn:2}
\end{gather}
where $\Psi_j(\zeta, \tau)$'s $(j=1,2)$ represent the complex-valued slowly varying field envelopes in the two channels of the $\mathcal{PT}$-symmetric couplers and $\kappa$ is the normalized inter-core linear coupling coefficient. In the above Eqs. (\ref{eqn:1}) and (\ref{eqn:2}) the second and third terms are, respectively, responsible for the group-velocity dispersion and self-phase modulation (cubic nonlinearity), whose coefficients are scaled to be one. Also, the terms proportional to $\Gamma$ in Eqs. (\ref{eqn:1}) and (\ref{eqn:2}), which make the system to be $\mathcal{PT}$-symmetric, are the balanced gain and loss, respectively. Note that in Eqs. (\ref{eqn:1}) and (\ref{eqn:2}), $\Gamma>0$ corresponds to type 1 (Fig. 1(a)) and $\Gamma<0$ refers to type 2 (Fig. 1(b)) $\mathcal{PT}$ coupler. Though these two simple configurations share the same notion of $\mathcal{PT}$-symmetry, in order to highlight which model among these two (gain/loss and loss/gain) exhibits rich dynamics, we have coined the above terminologies owing to the fact that the transmission properties may differ as the amplification and absorption are placed at different waveguides. In addition, we introduced these terms to avoid any confusion between the launching conditions, especially when the solitons are excited at the second waveguide which is assumed to be lossy in the typical $\mathcal{PT}$-symmetric dimer.

To identify the steering dynamics of solitons in the $\mathcal{PT}$-symmetric system, we first concentrate on the power-controlled steering, which is usually applied in all-optical steering following  a spatio-temporal evolution dynamics. In the former case, when the pump intensity is fixed at a low value, the device acts as a linear fiber coupler wherein one can observe that the evanescent coupling tends to steer the input pulse from channel 1 to channel 2. Nevertheless, at a high pump intensity, the nonlinearity is induced and owing to this effect, the energy of the input pulse steers back to the channel 1 itself. The system is said to be in the so-called unbroken $\mathcal{PT}$-symmetric regime as long as $\kappa>\Gamma$ \cite{ruter2010observation}. If the value of the gain/loss parameter equals the inter-core linear coupling coefficient, $\kappa$, the system results in a singularity condition (also known as \emph{exceptional point/super-symmetry}), whereas if the former exceeds the latter ($\kappa<\Gamma$), the  system is said to be in the broken $\mathcal{PT}$-symmetric regime \cite{BAMEPL2011}. In this work, we restrict the system to work in the $\mathcal{PT}$-symmetric unbroken regime rather than in the other domains. Hence we scale the inter-core linear coupling parameter, $\kappa$, and the gain/loss parameter, $\Gamma$, to be 1 and 0.5 respectively unless otherwise noted. The coupled Eqs. (\ref{eqn:1}) and (\ref{eqn:2}) are then numerically integrated by means of the pseudo-spectral method. To this end, we assume that the soliton profile with an amplitude $q$ is excited only in the first waveguide channel whereas the second channel is kept empty, i.e., $\Psi_1(\zeta=0, \tau)= q \sech(\tau)$,
$\Psi_2(\zeta=0, \tau)=0$, where $q^2$ indicates the pump intensity (input power). On the other hand, the transmission between the two channels is represented by a ratio of their corresponding integrated powers, as defined below: $T_j=P_j/(P_1+P_2)$, with $P(0) = \int_{-\infty}^{\infty} |\Psi_1(0,\tau)|^2 d\tau$, $P_j=\int_{-\infty}^{\infty} |\Psi_j(L_c,\tau)|^2 d\tau$, $(j=1,2)$.

\begin{figure}[t]
	\centering
	\includegraphics[width=0.4\linewidth]{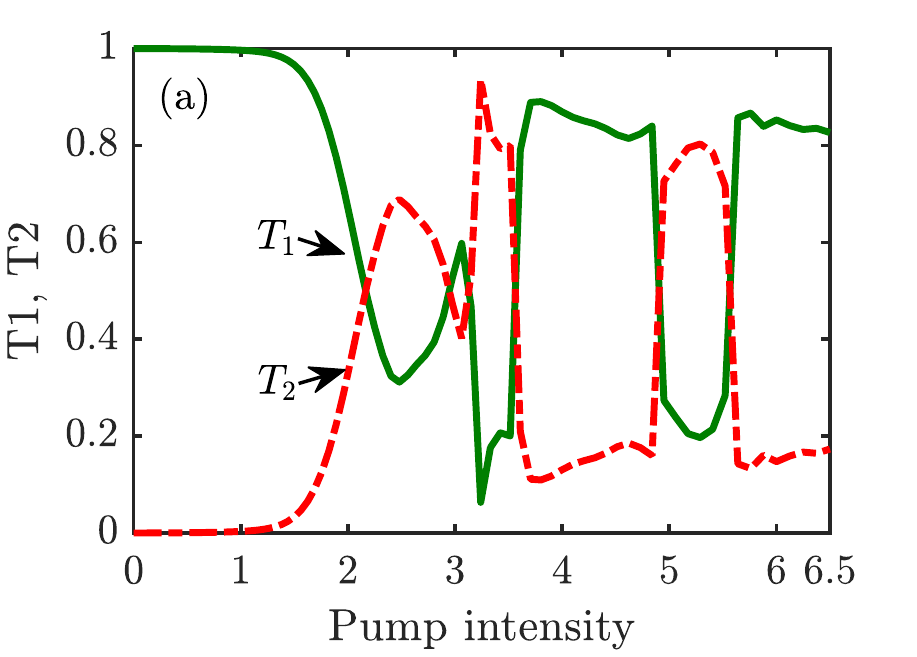}\hspace{0.2in}\includegraphics[width=0.4\linewidth]{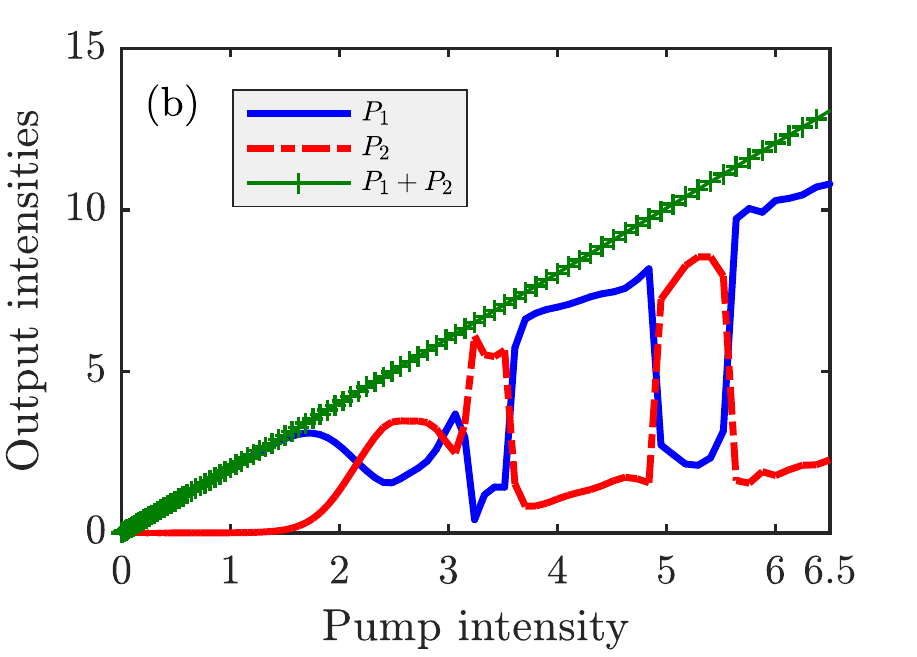}\\\includegraphics[width=0.4\linewidth]{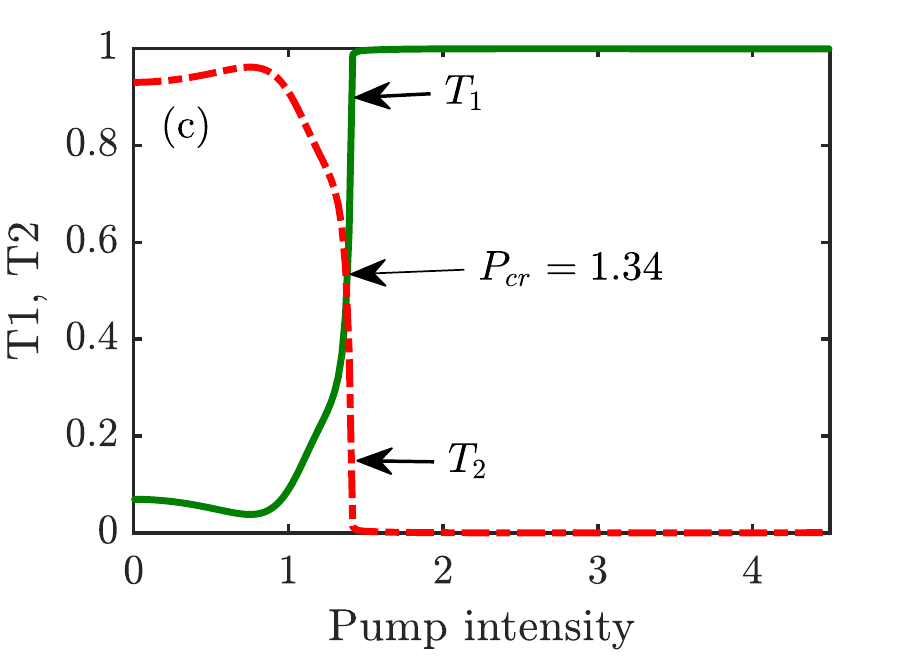}\hspace{0.2in}\includegraphics[width=0.36\linewidth]{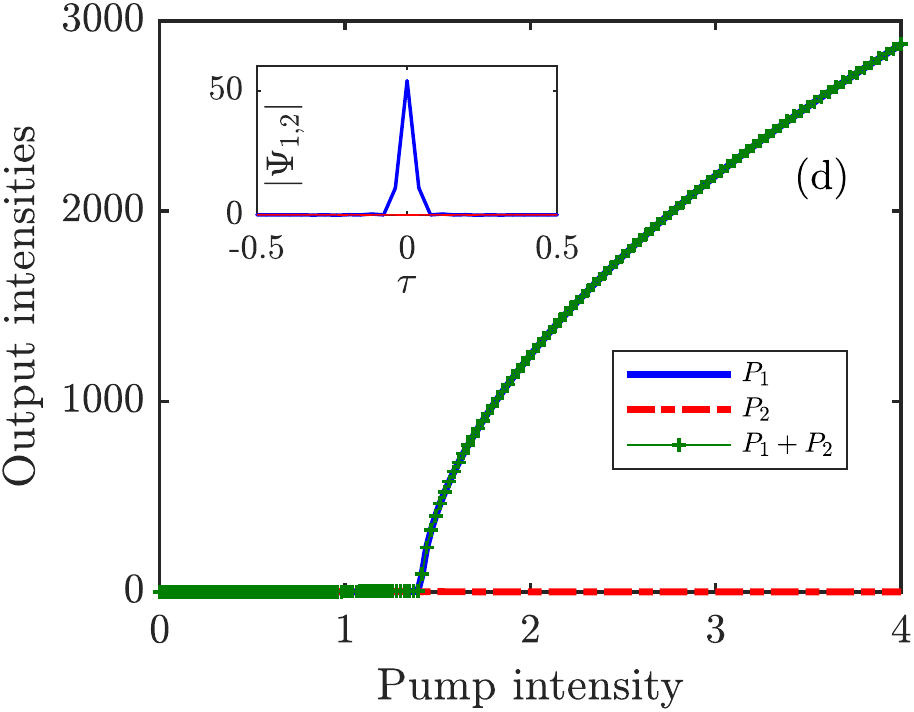}\\
	\topinset{\tiny(e)}{\includegraphics[width=0.45\linewidth]{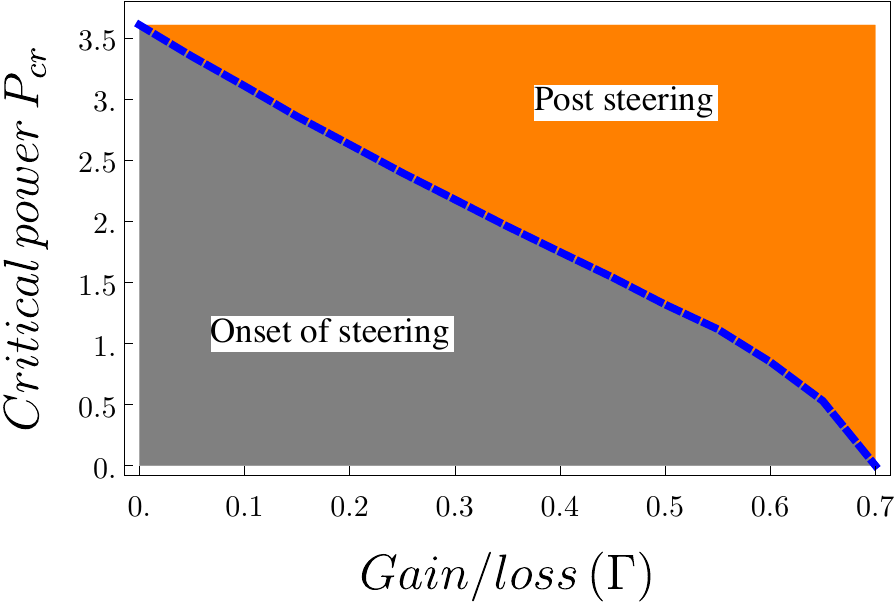}}{0.15in}{0.6in}\topinset{\tiny(f)}{\includegraphics[width=0.45\linewidth]{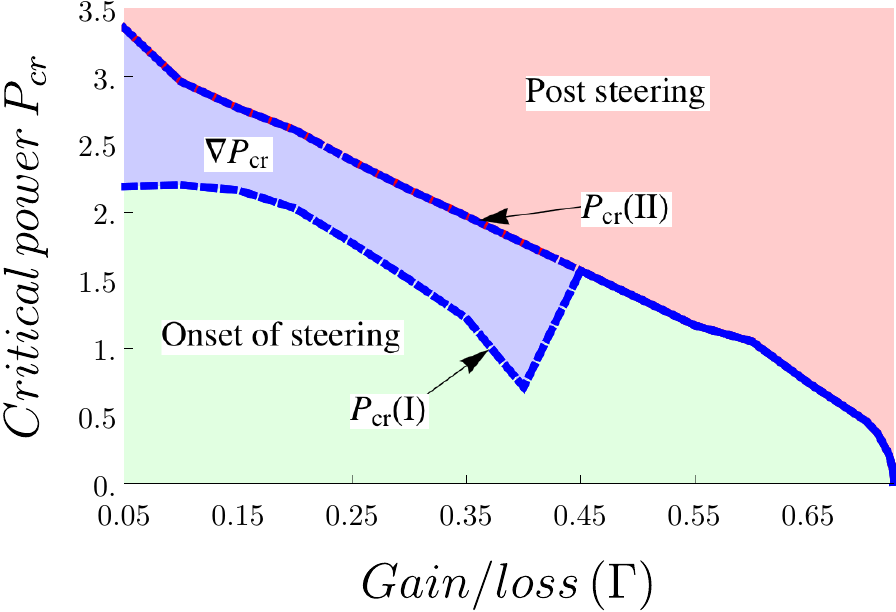}}{0.15in}{0.6in}
	\caption{The steering dynamics of solitons for $2\pi$ conventional couplers (top panels) and $2\pi$ $\mathcal{PT}$-symmetric couplers (middle panels), where inset in (d) demonstrates the amplified output pulse. Relation between the critical power and gain/loss parameter in ($\Gamma$, $P_{cr}$) plane for (e) $\pi/2$ and (f) $2\pi$ $\mathcal{PT}$ couplers.}
	\label{Fig2_2pi}
\end{figure}

Here $P(0)$ is the integrated power of the incident soliton pulse, and $P_1$ and $P_2$ are, respectively, the output powers of the transmitted pulses in the port 3 and port 4 of the two channels (Fig. \ref{FigSchematic}). Also, the parameter $L_c$ refers to the total coupling length of the system.
To demonstrate the soliton steering dynamics in the above system in detail, we considered couplers with different beating lengths ranging over $\pi/2$, $\pi$, $2\pi$ and $4\pi$. It is found that the $\mathcal{PT}$-symmetric effect  flawlessly works on $2\pi$ coupler, shown in Fig. \ref{Fig2_2pi} for type 1, contrary to the conventional couplers which generally exhibit richer steering dynamics only at the half-beating coupling length. viz. $\pi/2$ coupler \cite{trillo_switch,BAMPRE}. In the $2 \pi$ conventional coupler (see Figs. 2a and 2b), one can observe that there exists a multiple steering which is not periodic and non-smooth taking huge amount of pump power to steer. However, in the $\mathcal{PT}$-symmetric couplers, the inclusion of gain/loss parameter remarkably modifies the steering dynamics by reducing the critical power to a lower value ($P_{cr}=1.34$) compared to the conventional one. Further, the number of unwanted steering and oscillations observed above the first critical power in the conventional couplers have been completely canceled and thereby allowing one to achieve a very sharp steering and an excellent transmission efficiency, which stays unity (almost 99.99\% of energy transmission). Moreover, after the threshold intensity,  one observes that the output energy in channel 1 features extremely compressed (short) and intense pulse with a very few optical cycles (see Fig.2(d)). Indeed, the amplification is more than a factor of $10^3$ times compared to the one below the critical power. Also, the total transmitted energy is almost equal to that of the one found in the first waveguide, with negligibly very low energy shared in the channel 2, unlike the case with conventional couplers [cf. Figs. 2b and 2d].
\begin{figure}[t]
	\centering
		\topinset{\bfseries(a)}{\includegraphics[width=0.4\linewidth]{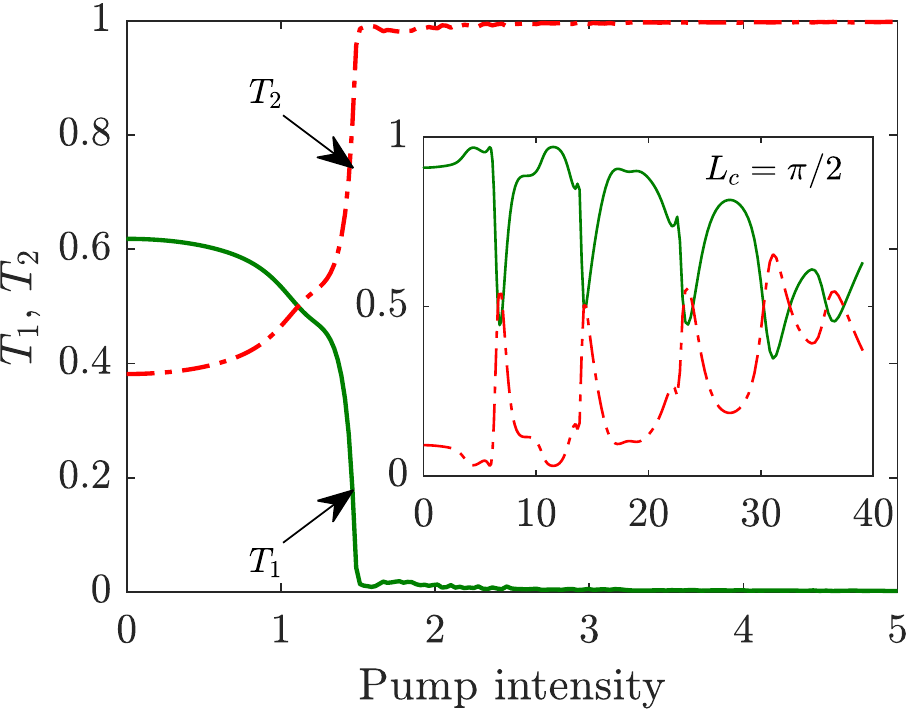}}{0.1in}{0.45in}\hspace{0.1in}\topinset{\bfseries(b)}{\includegraphics[width=0.45\linewidth]{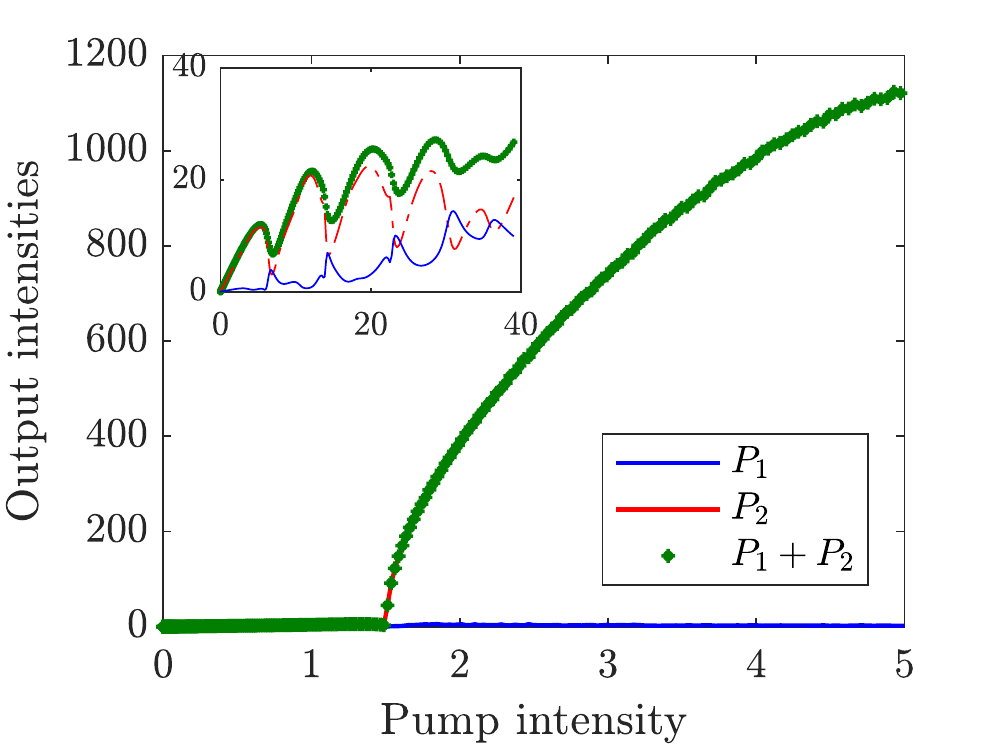}}{0.15in}{0.3in}\\
	\topinset{\bfseries(c)}{\includegraphics[width=0.45\linewidth]{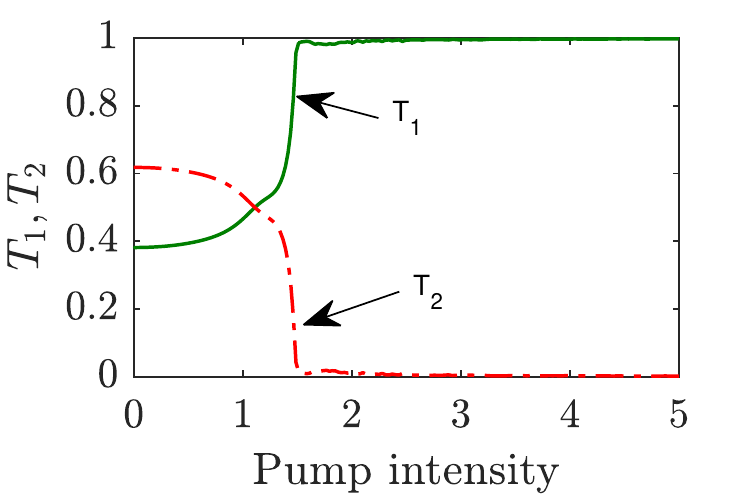}}{0.15in}{0.45in}\hspace{0.1in}\topinset{\bfseries(d)}{\includegraphics[width=0.45\linewidth]{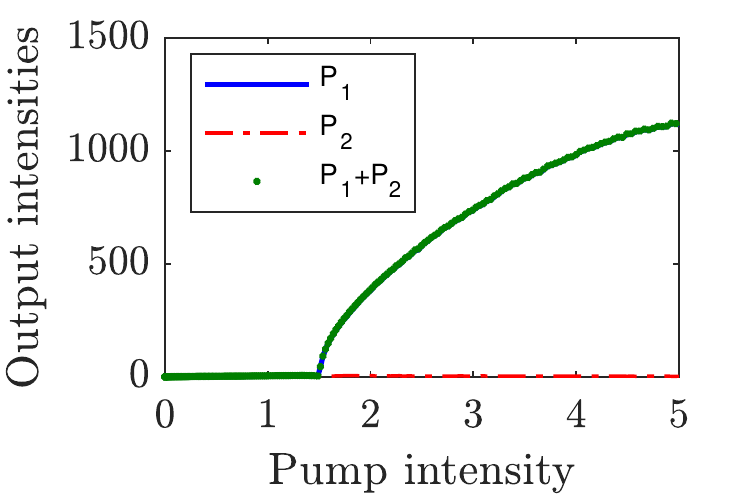}}{0.15in}{0.45in}
	\caption{(a, b) Steering dynamics in type 2 $\mathcal{PT}$ dimer of length $2\pi$ (inset $\pi/2$) and (c, d) different launching conditions in type 1 $\mathcal{PT}$ dimer with device length of $2\pi$ (Note halving of output intensity in Fig. 2(d)).}
	\label{Fig3_critical}
\end{figure}
\begin{figure}[!ht]
	\centering
	\topinset{\bfseries(a)}{\includegraphics[width=0.4\linewidth]{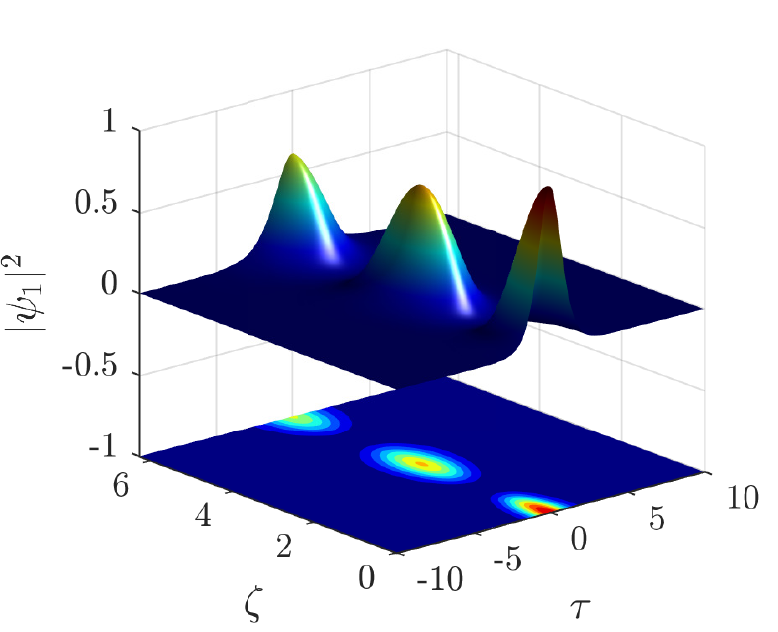}}{0.25in}{.4in}\hspace{0.15in}\topinset{\bfseries(b)}{\includegraphics[width=0.4\linewidth]{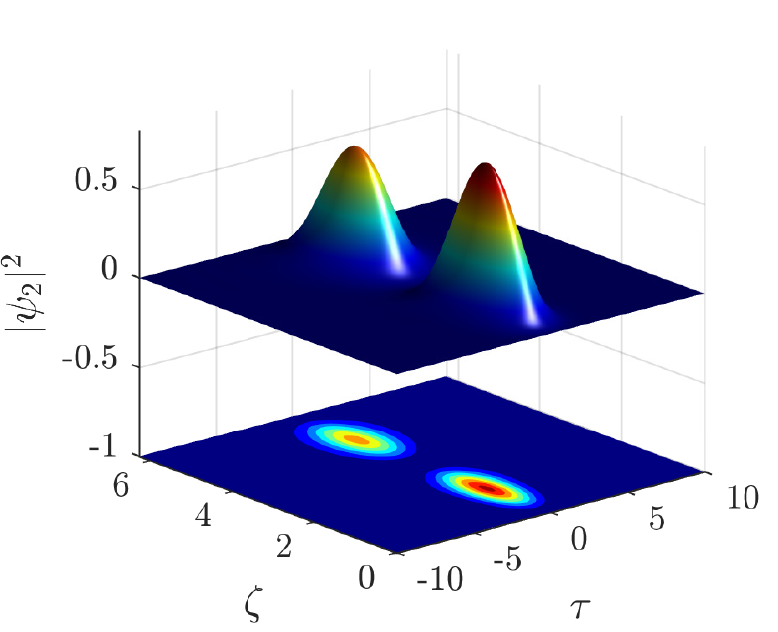}}{0.25in}{.4in}\\	\topinset{\bfseries(c)}{\includegraphics[width=0.4\linewidth]{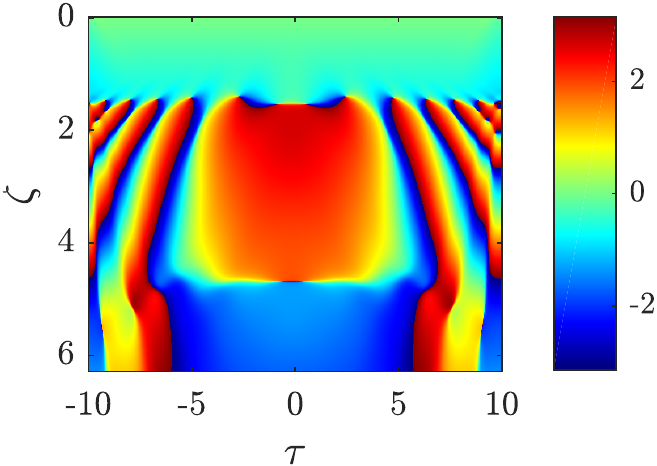}}{0.1in}{.25in}\hspace{0.15in}\topinset{\color{red}(d)}{\includegraphics[width=0.4\linewidth]{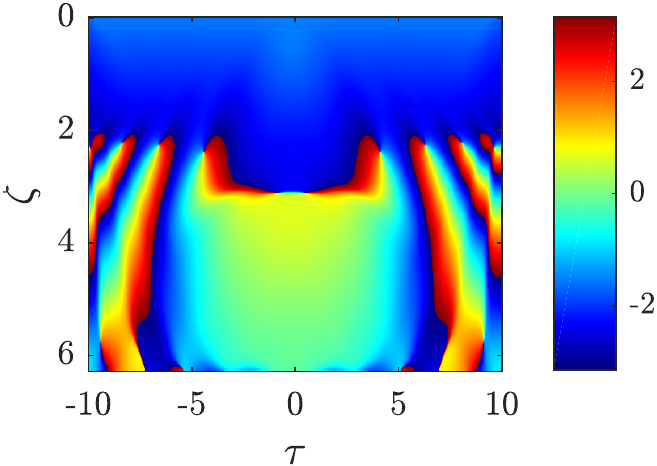}}{0.1in}{.25in}\\
	\topinset{\bfseries(e)}{\includegraphics[width=0.4\linewidth]{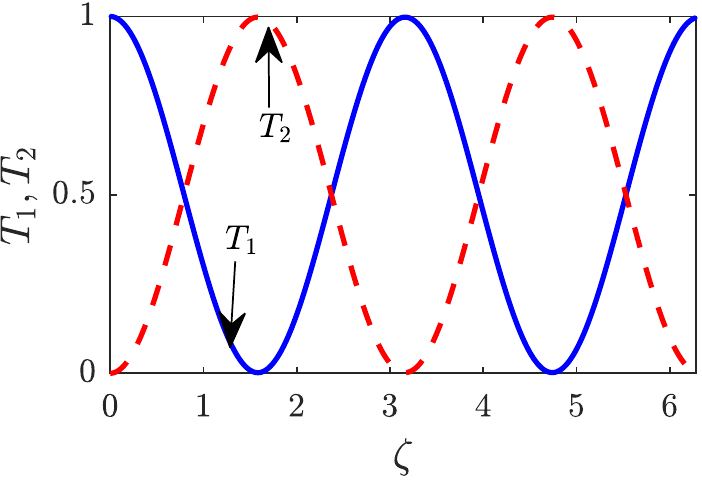}}{0.075in}{.38in}\hspace{0.175in}\topinset{\bfseries(f)}{\includegraphics[width=0.4\linewidth]{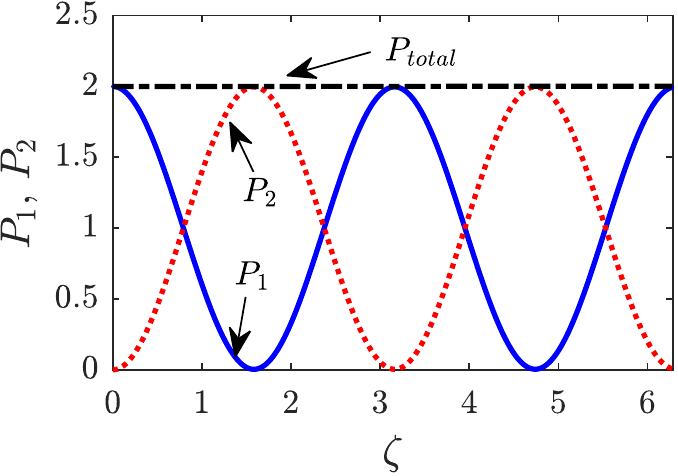}}{0.075in}{.35in}
	\caption{Top panels (a, b) depict the evolution of solitons in conventional couplers. Center panels (c, d) show the corresponding phase distributions and bottom panels reveal  (e) the total transmission and (f) output intensities as a function of propagation distance.}
	\label{Fig4_conventional}
\end{figure}
Further, Figs. \ref{Fig2_2pi}(e) and \ref{Fig2_2pi}(f) illustrate the role of the critical power in the $\pi/2$ and $2\pi$ $\mathcal{PT}$ dimers. It can be observed that in the $\pi/2$ $\mathcal{PT}$ dimer the critical power linearly decreases with an increase in the gain/loss parameter. In particular, when the gain/loss parameter approaches the singularity condition, it manifests as ultra-low power switch by yielding a critical power, $P_{cr}=0.035$, when the gain and loss parameter is  kept at $\Gamma=0.676$. Likewise, even in the $2\pi$ coupler the critical power drops to a lower value as the gain/loss parameter grows, however with a little different dynamics. Due to the inclusion of $\Gamma$, the number of threshold intensities is reduced to two for $\Gamma<0.45$. However, if we increase $\Gamma$ beyond $\Gamma>0.45$, it remarkably exhibits the sharp steering as observed in Fig. \ref{Fig2_2pi}(c). Also, it induces a phase shift between the two field components (cf. Figs. \ref{Fig2_2pi}(a) and \ref{Fig2_2pi}(c)). Further, here too quite remarkable ultra-low critical power is obtained as $P_{cr}=0.0107$ when $\Gamma$ is fixed at $\Gamma=0.7243$ (refer Fig. \ref{Fig2_2pi}(e)). Thus, from all the above ramifications, one can conclude that unlike the conventional coupler, which operates the steering dynamics fairly well only in the $\pi/2$ coupler, the $\mathcal{PT}$-symmetric coupler seemingly works for all domains of coupling lengths (in particular for the $2\pi$ coupler) with very low critical powers, which will firmly open up new possibilities for fabricating all-optical fiber components.
\begin{figure}[ht]
	\centering
	\topinset{(a)}{\includegraphics[width=0.4\linewidth]{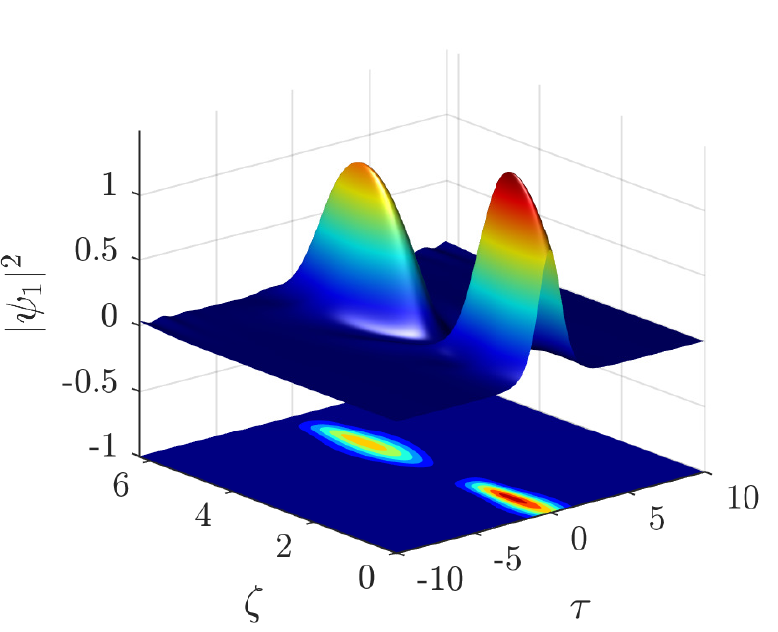}}{0.25in}{.4in}\hspace{0.15in}\topinset{(b)}{\includegraphics[width=0.4\linewidth]{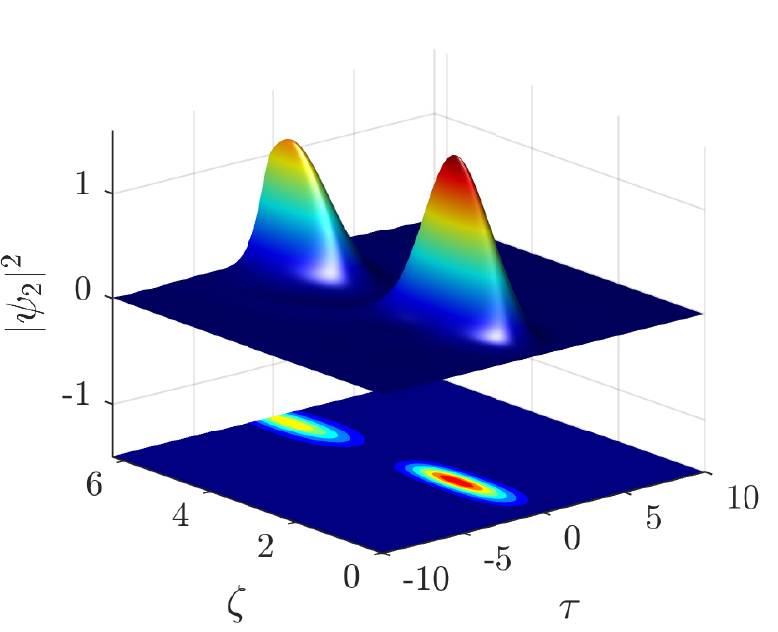}}{0.25in}{.4in}\\	\topinset{\bfseries(c)}{\includegraphics[width=0.4\linewidth]{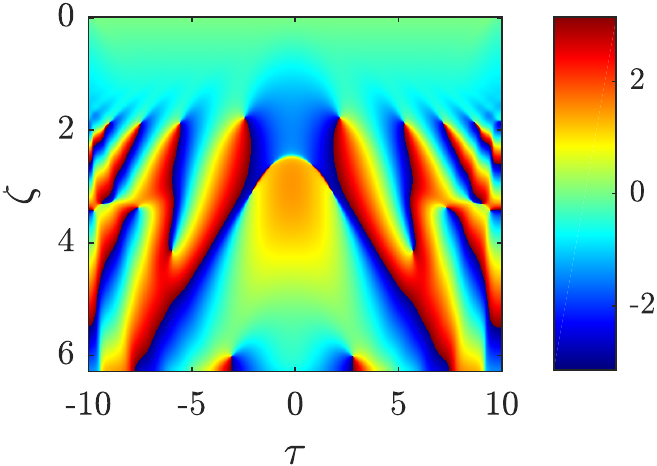}}{0.1in}{.25in}\hspace{0.15in}\topinset{\bfseries(d)}{\includegraphics[width=0.4\linewidth]{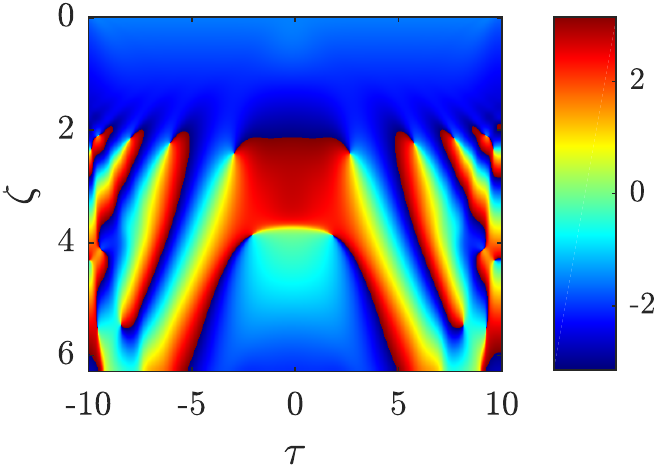}}{0.1in}{.25in}\\
	\topinset{\bfseries(e)}{\includegraphics[width=0.4\linewidth]{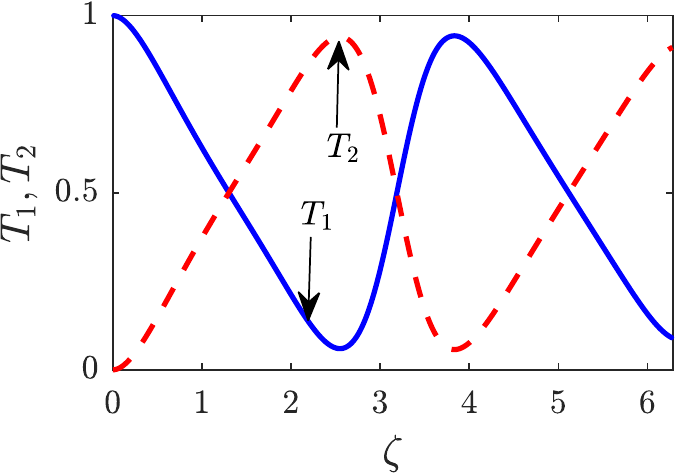}}{0.075in}{.45in}\hspace{0.175in}\topinset{\bfseries(f)}{\includegraphics[width=0.4\linewidth]{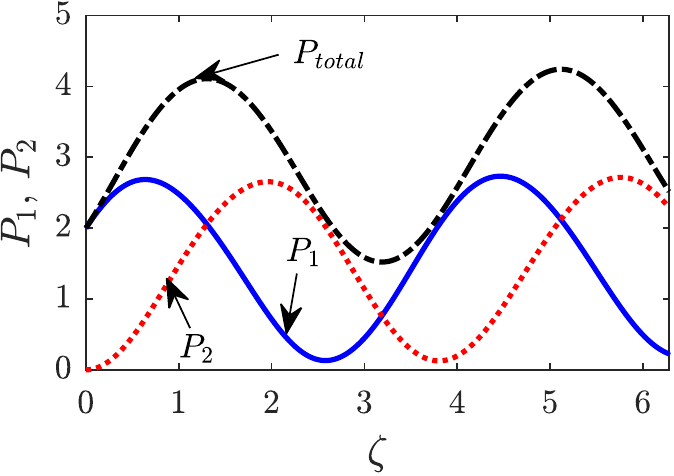}}{0.075in}{.3in}
	\caption{The same spatio-temporal dynamics of solitons in $2\pi$ $\mathcal{PT}$-symmetric couplers with type 1 $\mathcal{PT}$-symmetry configuration.}
	\label{Fig5_type1}
\end{figure}

Next, we discuss soliton steering in the following  alternative configuration of the $\mathcal{PT}$-symmetric coupler, with   the first waveguide having loss while the second one having equal amount of  gain (refer Fig. 1(b)). We call it as type 2 hereafter for convenience. As depicted in Fig. \ref{Fig3_critical}, it can be observed that the dynamics exhibited by the type 2 $\mathcal{PT}$-symmetric coupler is quite different from that of type 1. Here, unlike type 1, above the critical power, most of the energy is transferred to the cross state, thereby retaining the transmission efficiency high in the cross channel with a sharp steering.  It should be noted that this feature is enabled only if it is a $2\pi$ coupler, e.g. not in the $\pi/2$ coupler (see the inset of Fig. 3(b)).  On the other hand, in the $\pi/2$ coupler,  it is expected that a channel with gain will obviously exhibit amplification and will try to stay in the same medium as the nonlinear phase ($\phi_{NL}$) of the gain channel (no matter whether the gain is present in the bar or the cross channel) exceeds the lossy one by following the condition $\phi_{NL}>2\pi$ \cite{Agrawal2001}.  Owing to this fact, the steering dynamics is never complete and eventually it takes a huge pump intensity to finish the steering process.   But, in the type 2 $\mathcal{PT}$ coupler of length $2\pi$, it is totally different due to the delicate phase difference between  the gain and the lossy modes,  as well as the specific coupling length, which leads to  a new route of switching from  the input port to the cross channel. Thus the above two $\mathcal{PT}$-symmetric coupler  configurations enable two different routing mechanisms employing soliton pulses by merely swapping the gain and loss parameters. If we change the launching conditions in type 1 coupler (see Figs. 3(c) and 3(d),  it exhibits a bit of modification in the switching dynamics. Here we notice a small decrement in the transmission efficiency on the onset of steering (or in other words, it creates a 60:40 energy sharing in the linear domain like in type 2 $\mathcal{PT}$ coupler) and the amplification of output intensities is reduced to half compared to the former one (cf. Figs. 2(d) and {3(d)).}

In order to investigate the soliton propagation dynamics in the $\mathcal{PT}$-symmetric coupler,we depict the spatio-temporal evolution of the bright solitons in such a coupler. However, to understand the dynamics in proper perspective, we first demonstrate  spatio-temporal evolution of solitons in the conventional $2\pi$ coupler in Fig.\ref{Fig4_conventional}, where the soliton pulse is excited at the input port of channel 1. It may be observed that the pulse steers back and forth between the two channels and eventually exits at the output port of the same channel, which implies a route of the form $In_1\rightarrow Out_3$.  Moreover, the transmission of the pulses between the two channels is completely out of phase along the propagation direction (cf. Figs 4(c)--(4f)). But the $\mathcal{PT}$-symmetric type 1coupler (see Fig. \ref{Fig5_type1}) delivers a completely different sort of steering dynamics in sharp contrast to the conventional coupler. Here, the soliton pulse enters from input port of channel 1 and steers between the two arms and transmits at the output port of channel 2, where it makes a new energy route of $In_1 \rightarrow Out_4$. Compared to the conventional coupler, one can  easily observe that in both the output channels, the powers of the pulse get amplified and a different transmission route is noticed (see Figs. 5e and 5f). The possible mechanism for this new route  can be attributed to the fact that, in such $\mathcal{PT}$-symmetric couplers, the transmissions $T_1$ and $T_2$ are not in phase and there exists a phase shift of $\pi/2$ between them (refer Figs. 5c and 5d). However, in this type 1 $\mathcal{PT}$ coupler the total beating coupling length (analogous to the beat time period in quantum mechanics \cite{PRL.101.080402}) increases due to the influence of gain/loss parameter.
\begin{figure}[t]
	\centering
	\topinset{\bfseries(a)}{\includegraphics[width=0.4\linewidth]{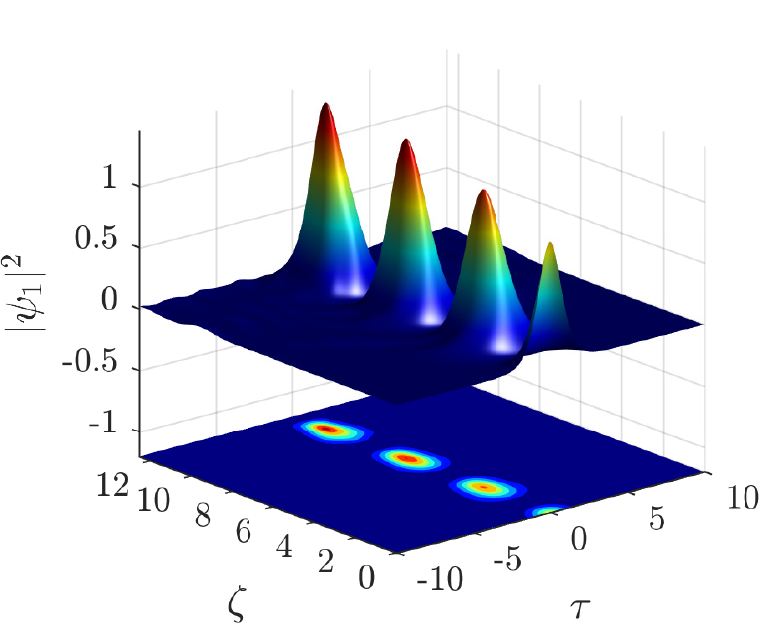}}{0.25in}{.4in}\hspace{0.15in}\topinset{\bfseries(b)}{\includegraphics[width=0.4\linewidth]{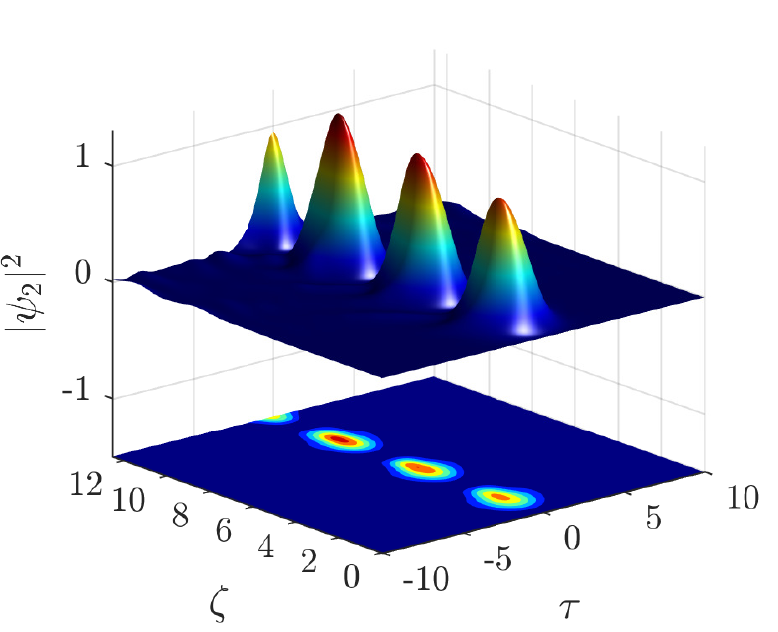}}{0.25in}{.4in}\\	\topinset{\bfseries(c)}{\includegraphics[width=0.4\linewidth]{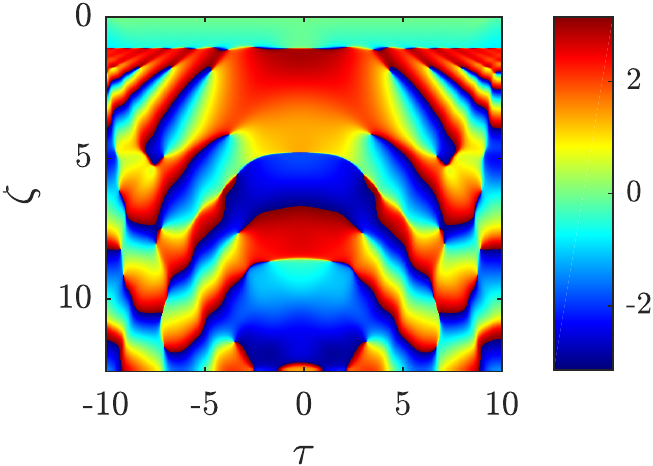}}{0.1in}{.25in}\hspace{0.15in}\topinset{\bfseries(d)}{\includegraphics[width=0.4\linewidth]{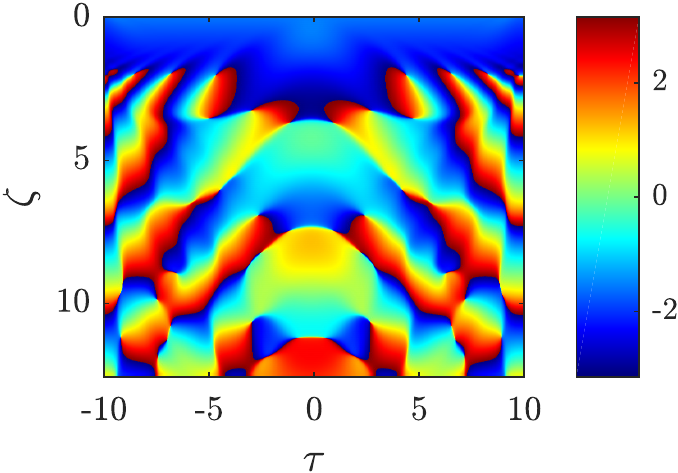}}{0.1in}{.25in}\\
	\topinset{\bfseries(e)}{\includegraphics[width=0.4\linewidth]{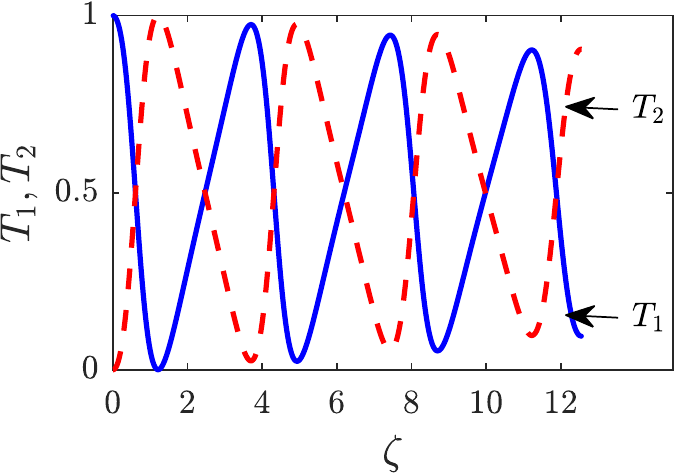}}{0.075in}{.55in}\hspace{0.175in}\topinset{\bfseries(f)}{\includegraphics[width=0.4\linewidth]{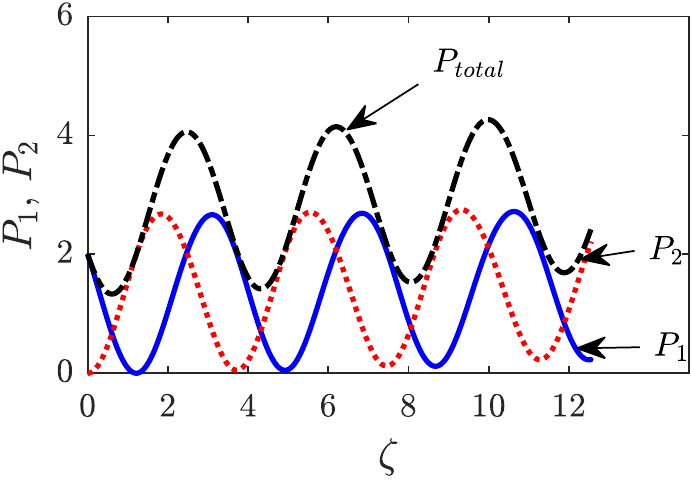}}{0.075in}{.45in}
	\caption{Evolution of optical solitons in type 2 $\mathcal{PT}$-symmetric couplers with the device length of $2\pi$.}
	\label{Fig6_type2}
\end{figure}

Next, we discuss the soliton propagation dynamics in type-2 $\mathcal{PT}$-symmetric coupler. As can be seen from Fig. \ref{Fig6_type2}, in this case, too, the soliton periodically steers to and fro between the two channels. If one increases the value of the loss/gain ($-\Gamma$) parameter, the following ramifications are observed. First, in both the channels, the pulse gets amplified along the propagation direction. Second, the energy of soliton pulses is equally shared between the two channels which opens up, possibly, a new $50:50$ energy sharing $\mathcal{PT}$-symmetric switch. Finally, the propagating soliton pulses are more stable in the two channels. Moreover, the coupling length of the steering dynamics is also dramatically decreased by causing a multiple steering in a given length of the coupler (see the bottom panels of Fig. \ref{Fig6_type2}). Overall, compared to the dynamics observed in type 1 $\mathcal{PT}$-symmetric coupler, the dynamics of the type 2 $\mathcal{PT}$-symmetric coupler delivers somewhat intriguing periodic steering than the former, for not only in the case of the $2 \pi$ coupler but also for other coupling lengths, with a very small energy loss in the form of radiation on the soliton background.

To realize the proposed system in practice, we suggest to adopt the typical parameters used in fiber optics communication systems. For instance, if we set $\beta_2=-20$ \si[per-mode=symbol]{ps^{2}\per\kilo\metre} near the wavelength $\lambda=1.55$ \si{\micro\metre}  and the pulse width $T_0=50$ \si{fs}, then dispersion length becomes $L_D=12.5$ \si{\centi\metre}. Hence the corresponding device length $L$  equals to  ($\zeta=z/L_D$), \SI{19.6}{\centi\metre} and \SI{78.5}{\centi\metre} for  $\pi/2$ and $2\pi$ couplers respectively. Also, consider the aforementioned critical powers as  $P_{cr}=0.035$ pertaining to $\pi/2$ and  $P_{cr}=0.0107$ for $2\pi$ couplers, which respectively give rise to \SI{28}{\watt}  and \SI{8.56}{\watt} in real units if we consider the nonlinearity as $\gamma=10$ \si[per-mode=symbol]{W^{-1}\per\kilo\metre}. However, when femtosecond solitons are used, higher-order effects such as third-order dispersion, shock terms and Raman shift may affect the switching operations of the $\mathcal{PT}$-symmetric dimer, which we leave for subsequent work.

To conclude, we have  studied the $\mathcal{PT}$-symmetric soliton switch in a system of coupled dimer with balanced gain and loss. We have shown that such $\mathcal{PT}$-symmetric soliton switch works remarkably well at a device length of $2\pi$, yielding an  ultra-low critical power of switching, which further paves the way for attainment of an ideal switch in a real-time (physical) system, even with the inclusion of loss.  We anticipate that the proposed soliton switch will open up possibilities for experimental realization and also revive the interest in all-optical soliton switching.

{\bf \large Funding.} Department of Science and Technology (DST) and Science and Engineering Research Board (SERB) (PDF/2016/002933 and SB/DF/04/2017, respectively).

{\bf \large Acknowledgment.} This research was also supported in part by the International Centre for Theoretical Sciences (ICTS) during a visit for participating in the program -- Non--Hermitian Physics --PHHQP XVIII (Code: ICTS/nhp2018/06).
\bibliographystyle{osajnl}
\bibliography{PT_Soliton}

\begin{thebibliography}{10}
\newcommand{\enquote}[1]{``#1''}

\bibitem{ruter2010observation}
C.~E. R{\"u}ter, K.~G. Makris, R.~El-Ganainy, D.~N. Christodoulides, M.~Segev,
  and D.~Kip, {\protect\JournalTitle{Nat. Phys.}} \textbf{6}, 192 (2010).

\bibitem{El-Ganainy2018}
R.~El-Ganainy, K.~G. Makris, M.~Khajavikhan, Z.~H. Musslimani, S.~Rotter, and
  D.~N. Christodoulides, {\protect\JournalTitle{Nat. Phys.}} \textbf{14}, 11
  (2018).

\bibitem{musslimani2008}
Z.~Musslimani, K.~G. Makris, R.~El-Ganainy, and D.~N. Christodoulides,
  {\protect\JournalTitle{Phys. Rev. Lett.}} \textbf{100}, 030402 (2008).

\bibitem{hodaei2014}
H.~Hodaei, M.-A. Miri, M.~Heinrich, D.~N. Christodoulides, and M.~Khajavikhan,
  {\protect\JournalTitle{Science}} \textbf{346}, 975 (2014).

\bibitem{SLonghi}
S.~Longhi, {\protect\JournalTitle{Phys. Rev. A}} \textbf{82}, 031801 (2010).

\bibitem{phang2014impact}
S.~Phang, A.~Vukovic, H.~Susanto, T.~M. Benson, and P.~Sewell,
  {\protect\JournalTitle{Optics letters}} \textbf{39}, 2603 (2014).

\bibitem{feng2013}
L.~Feng, Y.-L. Xu, W.~S. Fegadolli, M.-H. Lu, J.~E. Oliveira, V.~R. Almeida,
  Y.-F. Chen, and A.~Scherer, {\protect\JournalTitle{Nat. Mat.}} \textbf{12},
  108 (2013).

\bibitem{lin2011uni}
Z.~Lin, H.~Ramezani, T.~Eichelkraut, T.~Kottos, H.~Cao, and D.~N.
  Christodoulides, {\protect\JournalTitle{Phys. Rev. Lett.}} \textbf{106},
  213901 (2011).

\bibitem{suchkov2016nonlinear}
S.~V. Suchkov, A.~A. Sukhorukov, J.~Huang, S.~V. Dmitriev, C.~Lee, and Y.~S.
  Kivshar, {\protect\JournalTitle{Laser \& Photonics Reviews}} \textbf{10}, 177
  (2016).

\bibitem{PRA.94.023829}
S.~Karthiga, V.~K. Chandrasekar, M.~Senthilvelan, and M.~Lakshmanan,
  {\protect\JournalTitle{Phys. Rev. A}} \textbf{94}, 023829 (2016).

\bibitem{ramaswami2009}
R.~Ramaswami, K.~Sivarajan, and G.~Sasaki, \emph{Optical networks: a practical
  perspective} (Morgan Kaufmann, 2009).

\bibitem{chen1992}
Y.~Chen, A.~W. Snyder, and D.~N. Payne, {\protect\JournalTitle{IEEE journal of
  quantum electronics}} \textbf{28}, 239 (1992).

\bibitem{trillo_switch}
S.~Trillo, S.~Wabnitz, E.~Wright, and G.~Stegeman, {\protect\JournalTitle{Opt.
  Lett.}} \textbf{13}, 672 (1988).

\bibitem{kivshar1993switching}
Y.~S. Kivshar, {\protect\JournalTitle{Opt. Lett.}} \textbf{18}, 7 (1993).

\bibitem{driben2011}
R.~Driben and B.~A. Malomed, {\protect\JournalTitle{Opt. Lett.}} \textbf{36},
  4323 (2011).

\bibitem{Agrawal2001}
G.~Agrawal, \emph{Applications of nonlinear fibre optics} (Academic Press, New
  York, 2001).

\bibitem{BAMEPL2011}
R.~Driben and B.~A. Malomed, {\protect\JournalTitle{EPL (Europhys. Lett.)}}
  \textbf{96}, 51001 (2011).

\bibitem{BAMPRE}
I.~M. Uzunov, R.~Muschall, M.~G\"olles, Y.~S. Kivshar, B.~A. Malomed, and
  F.~Lederer, {\protect\JournalTitle{Phys. Rev. E}} \textbf{51}, 2527 (1995).

\bibitem{PRL.101.080402}
S.~Klaiman, U.~G\"unther, and N.~Moiseyev, {\protect\JournalTitle{Phys. Rev.
  Lett.}} \textbf{101}, 080402 (2008).

\end{thebibliography}
\end{document}